\documentclass[pdflatex,sn-mathphys-num]{sn-jnl}

\usepackage{graphicx}%
\usepackage{multirow}%
\usepackage{amsmath,amssymb,amsfonts}%
\usepackage{amsthm}%
\usepackage{mathrsfs}%
\usepackage[title]{appendix}%
\usepackage{xcolor}%
\usepackage{textcomp}%
\usepackage{manyfoot}%
\usepackage{booktabs}%
\usepackage{algorithm}%
\usepackage{algorithmicx}%
\usepackage{algpseudocode}%
\usepackage{listings}%

\theoremstyle{thmstyleone}%
\theoremstyle{thmstyletwo}%

\theoremstyle{thmstylethree}%

\begin{document}

\title[Article Title]{Improving GenIR Systems Based on User Feedback}


\author[1]{\fnm{Qingyao} \sur{Ai}}\email{aiqy@tsinghua.edu.cn}
\equalcont{These authors contributed equally to this work.}

\author[2]{\fnm{Zhicheng} \sur{Dou}}\email{dou@ruc.edu.cn}
\equalcont{These authors contributed equally to this work.}

\author*[1]{\fnm{Min} \sur{Zhang}}\email{z-m@tsinghua.edu.cn}

\affil*[1]{\orgdiv{Dept. of Computer Science and Technology}, \orgname{Tsinghua University}, \city{Beijing}, \country{China}}

\affil[2]{\orgdiv{Gaoling School of Artificial Intelligence}, \orgname{Renmin University of China}, \state{Beijing}, \country{
China}}


\abstract{In this chapter, we discuss how to improve the GenIR systems based on user feedback. Before describing the approaches, it is necessary to be aware that the concept of "user" has been extended in the interactions with the GenIR systems. Different types of feedback information and strategies are also provided. Then the alignment techniques are highlighted in terms of objectives and methods. Following this, various ways of learning from user feedback in GenIR are presented, including continual learning, learning and ranking in the conversational context, and prompt learning. Through this comprehensive exploration, it becomes evident that innovative techniques are being proposed beyond traditional methods of utilizing user feedback, and contribute significantly to the evolution of GenIR in the new era. We also summarize some challenging topics and future directions that require further investigation.}

\maketitle

\section{Introduction}\label{intro}
For an information access system that is built to provide useful information to the users, interactions with users are definitely crucial and important. There are two types of user feedback: \textit{explicit feedback} and \textit{implicit feedback}, based on whether the user's opinions or preferences regarding the provided information are expressed directly through clear statement or indirectly with some signals. (Section 1.2 will delve into more details about explicit and implicit feedback.) 

Similar to traditional information retrieval, generative information retrieval (GenIR) also leverages users’ feedback feedback in various ways to improve the system’s capability and performance. However, upon closer examination, it becomes evident that there are distinct factors that specifically contribute to GenIR systems, in terms of types of feedback information and utilization strategies. We also raise the attention to the definition of “user” in the new era. 

In this chapter, we first discuss the feedback factors and the differences in the new era in the first section. Subsequently, we provide detailed descriptions and discussions on the alignment with user factors in GenIR in Section~\ref{align}.  “Alignment” in generative models usually refers to the process of fine-tuning the model to ensure its generated text aligns with specific goals, values, or user intentions, often through human feedback or instruction fine-tuning. Subsequent to this, Section~\ref{sec:userfeedback} explores the user feedback learning in GenIR, which involves training and improving the GenIR system through understanding users' intents, interests, or preferences based on their historical feedback. A summary is given in the final section, Section~\ref{summary}, along with discussions on the challenges and future directions. 

\subsection{Concept of User in GenIR Era}
Over past years, discussions about the \textbf{\textit{user}} in information access systems, including search engines, recommender systems, question-answering platforms, etc, have primarily centered around human beings interacting with these systems. However, in the emerging GenIR era, where the new IR system is designed to connect with human beings, tools, or even other GenIR systems, the concept of the \textbf{\textit{user}} has been enlarged to a much broader sense. 

A \textbf{\textit{user}} of the GenIR system can now include:

\begin{itemize}
    \item A human being who uses the GenIR system, similar to the \textit{user} in traditional IR system;
    \item A Large Language Model (LLM) agent that can send or receive information to or from the GenIR system, or engage in bi-directional information exchange, also refers to the \textit{agent} in publications;
    \item Another system, tool, or application that interacts with the GenIR system, sometimes termed as the \textit{client} in technical context.
\end{itemize}
Interactions from the traditional \textit{users}, \textit{agents}, and \textit{clients} should all be taken into consideration as user feedback, whether it from real or virtual \textbf{\textit{users}} within the GenIR system.

\subsection{User Feedback} 
The GenIR system still maintains two fundamental types of user feedback: \textit{implicit feedback} and \textit{explicit feedback} as usual. However, the scope of feedback information has been significantly broadened.

Consistently, one of the major feedback is the user interaction history. The interactions with the GenIR system encompass queries, questions, clicks, views, purchases, comments, and more. Such information is usually taken as \textit{implicit feedback} information. In contrast, users’ explicit annotations such as favorites, likes, ratings, or direct feedback on satisfaction constitute \textit{explicit feedback} information. In GenIR systems, a notable difference lies in the increased availability of \textit{explicit feedback} provided by users through system inputs or prompts. Nowadays, users are accustomed to communicating their specific requests, intentions, and interests to the GenIR system. In many instances, multi-round conversations have become commonplace.

In the GenIR system, feedback information manifests in two distinct forms: 

1) Numerical information, primarily consisting of ID-level data that identifies the items with which the user has interacted. Sometimes, this information is presented in sequential order. Such information offers a glimpse into the user's behavioral patterns. 

2) Detailed information encompassing multiple modalities. Textual data is the most commonly utilized, including query text, item titles, content, user comments, and questions. As LLM technology rapidly advances, longer natural language expressions are increasingly being leveraged. Multimedia inputs, such as images, music, or videos, sometimes integrated with visal LLM (e.g.~\cite{liu2024rec}), have also garnered significant attention. This multifaceted feedback allows for a richer, more nuanced understanding of the user's preferences and interactions within the GenIR system.

How to leverage such rich user feedback information smoothly in the GenIR system to improve performance is a crucial part in the new LLM era. The next section briefly summarizes the strategies for using user feedback information in the GenIR system.

\subsection{Strategies for GenIR System Improvement with User Feedback}

Introducing user feedback information into the GenIR system can be facilitated through prompt engineering or instruction construction, which is perhaps the most straightforward approach~\cite{liu2024rec, dai2023uncovering, liu2023chatgpt,wang2023zero}. Historical user interactions can be encoded using various types of index, such as title-based indexing, random indexing, independent indexing, sequential indexing, semantic indexing, or collaborative indexing~\cite{geng2022recommendation}. These indexing inputs can then serve as prompts for the system, as illustrated in the following Figure \ref{fig:input-index}. The prompt strategy is commonly employed in zero-shot or in-context learning scenarios, where LLMs are directly utilized as the information system. By leveraging this approach, the GenIR system can efficiently integrate user feedback to enhance its performance.

\begin{figure}
	\centering
	\includegraphics[width=1.0\linewidth]{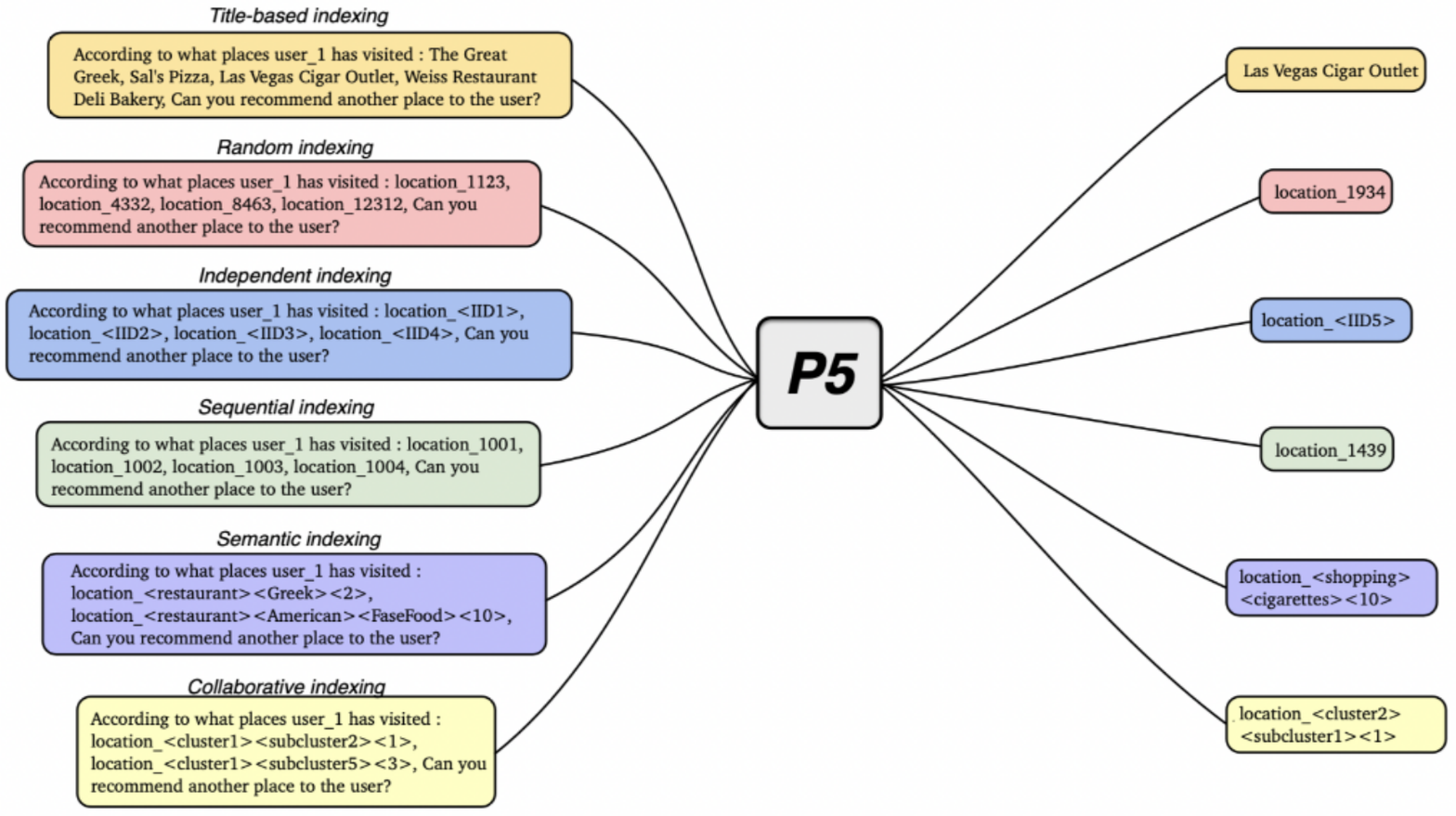}
	\caption{Different types of indexing as prompt input to GenIR systems proposed by \citet{geng2022recommendation}.}
	\label{fig:input-index}
\end{figure}

The second strategy involves leveraging historical interaction information for fine-tuning the parameters of the LLM. This approach aligns the user or item representations within the pre-trained language model. Typically, this information is utilized as either an ID index or a text index with item content, reviews, etc~\cite{hou2024large, rajput2024recommender, zhai2023knowledge}. In certain instances, the user-item collaborative information is initially encoded by a traditional IR system to generate embeddings for users or items~\cite{liao2023llara, luo2024integrating, petrov2023generative}. These embeddings implicitly contain collaborative interaction feedback, which is subsequently utilized in the subsequent fine-tuning and alignment process.

The third strategy focuses on capturing the user's implicit or explicit preferences, and indicating both vague and specific intents. By integrating these preferences and intents, the GenIR system is able to identify the user's specific task~\cite{zhang2023recommendation}. For instance, an implicit preference associated with a specific search intent for a mobile phone would be linked to a product-search task. While an implicit preference with a more vague intent might lead to a recommendation task.

The fourth strategy is to take user behavior as the action, reward, or even the evaluator within an Agent-based GenIR system~\cite{zhang2023generative, huang2023recommender,shu2023rah,wang2023recagent,wang2023recmind}. This approach effectively guides the system in learning the appropriate actions and responses. In such GenIR systems, user feedback information plays a crucial role in the system's learning and refinement process across multiple rounds of interaction. By continuously incorporating user feedback, the system can adapt and improve its performance.

It is anticipated that even more strategies will emerge as research continues. In the subsequent sections, we discuss deeper into these various approaches, exploring the alignment with user preferences and the learning mechanism.

\section{Alignment with User Factor in GenIR }\label{align}









Alignment techniques have been widely recognized as one of the key components for the construction of effective large language models (LLMs). 
In the first technical report of ChatGPT~\cite{openai2023gpt4}, alignment techniques such as Reinforcement Learning with Human Feedback (RLHF)~\cite{NIPS2017_RLHP,bai2022training} has already been extensively used in the training and construction of the chat system. 
Right after the success of ChatGPT, LLM alignment has become one of the most important research directions in the community of natural language processing, and, as discussed in the later part of this chapter, also has significant potential for building information access systems in the era of generative AI. 

Despite the recent surge in interest in alignment technology following the success of ChatGPT, it is important to note that research in this area has been ongoing for many years. 
In fact, it's difficult to pinpoint the exact moment when alignment techniques were first introduced to the studies of large language models.
In the early days of neural language models, researchers focused mainly on designing more powerful model structures~\cite{NIPS2013_word2vec,rnn,NIPS2017_attention} and training techniques~\cite{peters2017semisupervised,devlin2019bert} to enable LLMs to process information and learn patterns from massive amounts of data.
The performance of language models in those days were not strong enough to understand user's instructions and generalize to multiple types of tasks.
However, as the capability of LLMs grows, concerns related to aspects other than task performance, such as ethic, robustness, and etc., have gradually become obstacles that prevent LLMs from applications in real-world scenarios.
For example, in 2016, users had successfully tricked a twitter bot constructed by Micorsoft (i.e., Tay) to produce statements that were misogynistic~\cite{Vincent2016}; in 2022, Meta's BlenderBot 3 had been ``taught'' to be raciest right after it was released to the public~\cite{Silva2022}. 
Therefore, a greater emphasis, especially on model safety and ethical issues, was placed on aligning LLM outputs with human values and preferences since 2017.
It was around this period when a group of alignment methods, including the famous RLHF, have been used in the training of LLMs. 
As of today, almost all LLMs must go through an alignment process before being launched and released to the public. 

In this section, we forus on the introduction and discussion of LLM alignment from the perspectives of information access. 
Besides the safety and ethical problems of LLMs, there are also uniques challenges and needs of LLMs when applied to information access tasks. 
Those unique challenges also lead to unique methods and research directions that have great potential for information accessing in the era of generative AI.
In the followings, we first provide a brief introduction of the common objectives for alignments in LLMs and information accessing, and then introduce a couple of representative alignment methods in the field. 
Last but not least, we discuss the connections and differences between alignments and other LLM techniques such as supervised fine-tuning (SFT) from the perspective of information retrieval and access.

\subsection{Alignment Objective}\label{sec:align_objective}

LLM alignment is a cornerstone in the development of generative AI systems, particularly in the context of ensuring that these models act in ways that are beneficial, safe, and aligned with human values and intentions. 
One the one hand, as LLMs become more powerful everyday, their potential impact on human society increases, making the alignment of these models with ethical standards and user intentions an essential objective~\cite{bai2022training,zhang2023safetybench,openai2023gpt4}.
On the other hand, LLM alignment techniques can supplement supervised fine-tuning (SFT) or other training techniques in equipping LLMs with abilities or characteristics desired for diverse tasks and applications in specific domains~\cite{NEURIPS2022_instruction}. 
From the perspectives of information accessing, the objective of LLM alignment techniques is multifaceted, with some parts of it aligning closely with other LLM applications and some parts of it diverges significantly from those widely considered in the development of general LLMs.   

\subsubsection{Objectives Shared by General LLM Applications}
Similar to other LLM applications, the usage of LLMs in information accessing needs to prevent the possibility of outputs that are harmful from ethical perspectives or undesirable by user intents.
Specifically, such objectives include but not limited to:

\textbf{Preventing Harmful Outputs}.
One primary objective of LLM alignment techniques, shared by both information accessing and other applications, is to prevent models from generating harmful, biased, or inappropriate content~\cite{bai2022training} that violates the universal values of human beings. 
This includes outputs that could be misleading, factually incorrect, or that perpetuate harmful stereotypes. 
Before the era of generative AI, major information accessing systems usually focus on retrieving existing web pages or documents created by human to satisfy user's information need. 
Therefore, the prevention of harmful outputs can be done directly through pre-processing such as spam detection and keyword filtering~\cite{10.1145/2207243.2207252,10.1145/1099554.1099671}. 
With LLMs, however, controlling the outputs of information systems becomes significantly more difficult due to their stochastic nature~\cite{wolf2024fundamental}. 
Alignment techniques that prevent such harmful outputs through the post-training of LLMs have then become the most popular methods used in generative systems.

\textbf{Aligning with User Intents}.
Another critical aspect of LLM alignment is ensuring that models accurately understand and align with user intentions~\cite{NEURIPS2022_instruction}. 
This means that LLMs must be adept at interpreting the context and nuances of user queries and generating responses that accurately reflect the user's desired outcome. 
User intent understanding is at the core of information accessing, and numerous methods have been proposed to solve this problem in the context of traditional matching-based IA systems~\cite{10.1145/1772690.1772714,10.1007/978-3-642-00958-7_53,10.1145/3159652.3159714}
Unfortunately, as the internal knowledge structure of LLMs are still obscure, it's difficult (at least of today) to adopt our knowledge and experience obtained from previous studies directly to generative IA systems. 
LLM alignment is one of the most direct and practical methods to improve the system's ability in understanding user intents.

\textbf{Adhering to Ethical Guidelines}.
LLM alignment also involves adhering to ethical guidelines and principles. 
This encompasses a range of considerations, from ensuring privacy and data security to promoting fairness and avoiding discrimination.
In information accessing, these are also of great importance in practice. 
Popular search engines before the era of generative AI have already been widely criticized for imposing biased exposure to information such as political statements and news~\cite{doi:10.1080/1369118X.2020.1764605,doi:10.1073/pnas.1419828112}.
With more powerful yet nontransparent LLMs used in modern IA systems, such issues are becoming more intricate and vital. 
Developing generative IA systems that can balance fairness with relevance, respect user privacy, and treat sensitive topics with the appropriate level of care requires a deliberate and thoughtful approach. 
More importantly, as cultures, individuals, and groups may have vastly different views on what is considered appropriate, ethical, or aligned with their intentions, we need methods that are both effective and efficient in terms of model adaption.

\subsubsection{Objectives Unique to Information Accessing}

Besides those common alignment objectives shared by general LLMs, information accessing also has unique challenges that must be solved in order to construct effective generative IA systems.
As the ultimate goal of information retrieval and access is to satisfy user's information needs, to the best of our knowledge, the special characteristics needed by generative IA systems can be broadly categories into two types in existing literature: the need of personalization, and the capability of fine-grained information discrimination\footnote{Please note that the categorization here is by no means inclusive as this is still an ongoing research topic.}.

\textbf{Personalization}.
At its core, personalization is about tailoring the interactions and information delivery of a system to the unique preferences, interests, and needs of an individual user~\cite{10.1145/1721831.1721835}. 
The key idea is to transfer the user experience from a one-size-fits-all approach to a more intimate and relevant exchange~\cite{10.1145/1321440.1321515,10.1145/2348283.2348312}. 
In general applications of LLMs, this usually means speaking with the languages, styles, and values preferred by each user.
In the context of information accessing, this also means understanding and utilizing the connection between user's information need with time, locations, application scenarios, and all kinds of user information that potentially affect user's perceptions of information utility. 
Traditional personalization in information accessing focus on the construction of user profiles and the design of algorithms and models that effectively incorporate user information into the analysis of information relevance. 
In the era of generative AI, while the structures of the models and systems have significantly changed, the needs of those two still exist.
LLMs have strong in-context learning ability, which can implicitly create a user model simply by feeding the descriptions of user profile as prompts in the input user queries.
Yet, existing LLMs can only take inputs with text, images, or other standard multimedia formats, but user profiles go beyond these. 
How to construct and incorporate hyper-information like user-user, user-item, and item-item interactions effectively under the current model frameworks of LLMs and generative AI is still an open question.
Alignment techniques, as flexible and relatively lightweight methods to optimize large generative models, are thus of great potentials for personalization in generative information accessing.

\begin{figure}
	\centering
	\includegraphics[width=0.8\linewidth]{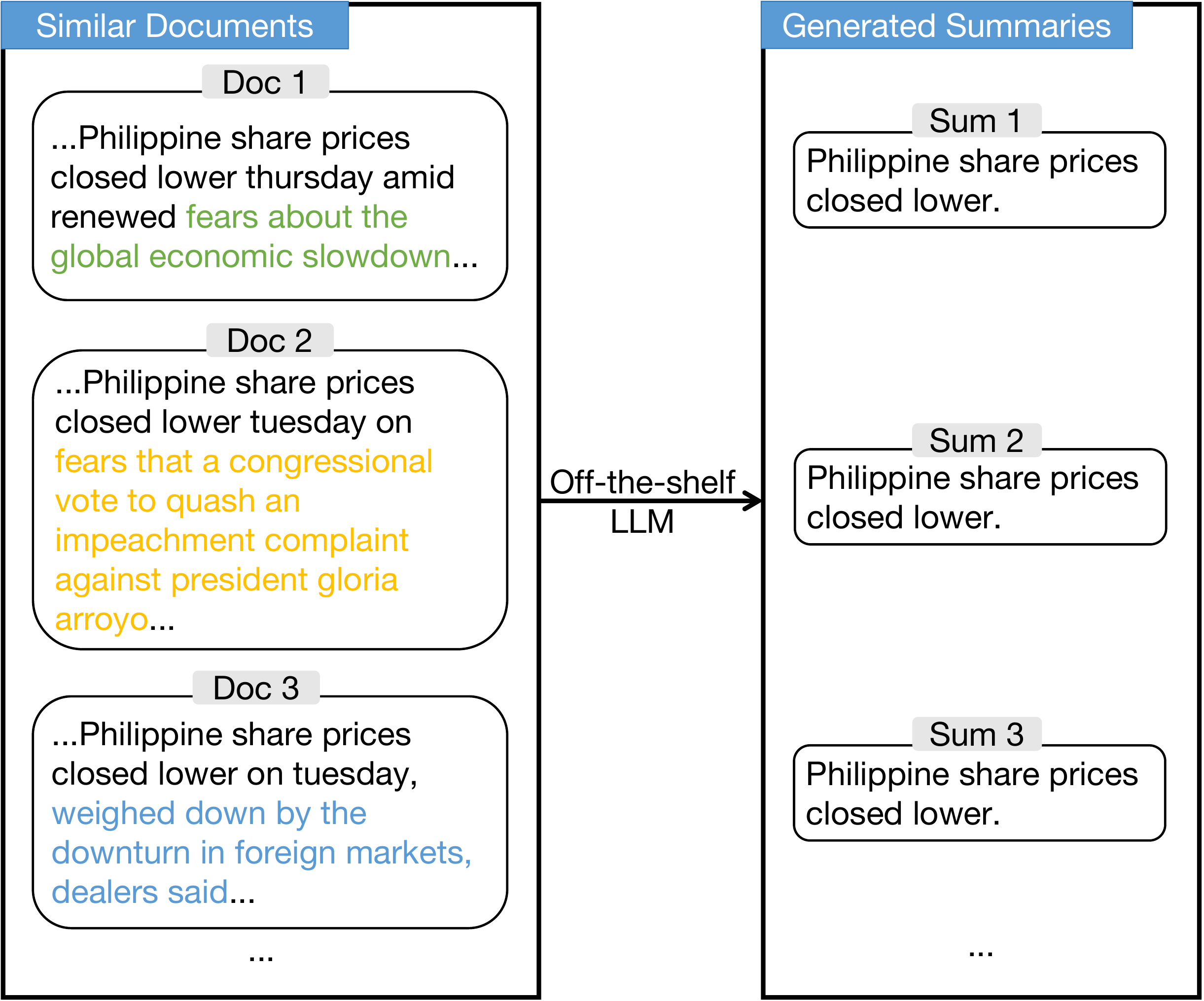}
	\caption{Illustrations of LLMs application in document summarization for similar documents provided by \citet{dong2023aligning}. The distinctive parts of each document are highlighted in different colors.}
	\label{fig:llminir}
\end{figure}

\textbf{Fine-grained Discrimination}.
With the exponential increase in the availability of digital content, the challenge of information access is no longer just finding relevant information but finding the most appropriate content among a set of potentially relevant items.
Most existing studies on retrieval and ranking models are essentially developing better methods to analyze and discriminate input documents based on their fine-grained differences and utility given user's queries.
In the era of LLM, while the final outputs of an IA system may no longer be a simple listing of result candidates, the ability of discriminating information in fine grains is still of great importance.
It allows the system to navigate in complex data collections, identifying the subtleties that differentiate pieces of information in ways that are significant to the user. 
Yet, the acquisition of such ability is usually not covered in the alignment process of off-the-shelf LLMs. 
An illustration example is provided by~\citet{dong2023aligning} in Figure~\ref{fig:llminir}.
When the request is the same (i.e., ``create a summary of the document'' in Figure~\ref{fig:llminir}) and the input documents are similar, the off-the-shelf LLM (i.e., Flan-T5 in the figure) tend to produce identical responses to all documents. 
Such problems could be insignificant in many NLP applications where the quality of outputs is evaluated independently with each other. 
In information accessing, however, we often care about the discrimination of input documents more than we care about their absolute relevance or utility. 
For instance, if we generate identical snippets for similar documents retrieved by search engines, it would remarkably increase the difficulty for users in pinpointing the exact result that answer her needs.
Because the ability to produce such discriminative outputs in fine grains can hardly be learned from the standard next token prediction tasks, one may need extra alignment process to enhance the model from this perspective~\cite{yoon2024ask}.

\subsection{Alignment Method}

Alignment methods are strategies developed to steer the behavior of generative AI models towards desired outcomes. 
These methods often rely on a framework that involves computing rewards based on model outputs and using these rewards to optimize the model's performance. 
Specifically, this usually involves two steps: (1) the collection and computation of rewards, and (2) the optimization of model parameters based on the rewards.
Most existing alignment methods are designed for general alignment objectives in NLP tasks, but, given the great potential and importance of LLMs in future information access, several researchers have also started to investigate how to design alignment methods tailored for the needs of IA tasks.
In this section, we first introduce a couple of well-developed reward collection methods in LLM alignments, and then briefly describe several standard optimization methods that have been widely used in existing studies.

\subsubsection{Collection of Rewards}\label{sec:reward}


\begin{table}[t]
	\centering 
	\caption{Example alignment methods and their categorization based on input and computation/training paradigms.}
	\label{tab:alignment_types}
	\begin{tabular}{lcccc}
		\toprule
		& \multicolumn{2}{c}{Input} & \multicolumn{2}{c}{Computation/Training} \\ 
		\cmidrule(lr){2-3} \cmidrule(lr){4-5}
		& Pointwise &  Pairwise/Groupwise & Pointwise & Pairwise/Groupwise \\ 
		\midrule
		RLHF & \cite{NIPS2017_RLHP,bai2022training,ziegler2020finetuning} & & \cite{bai2022training} &  \cite{NIPS2017_RLHP,bai2022training,ziegler2020finetuning}\\ 
		RLAIF & \cite{lee2023rlaif,yang2023rlcd} & & \cite{yang2023rlcd}&\cite{lee2023rlaif,yang2023rlcd}  \\ 
		RLCF & & \cite{dong2023aligning} & \cite{dong2023aligning} &  \\ 
		\bottomrule
	\end{tabular}
\end{table}

\begin{figure}[t] 
	\centering
	\includegraphics[width=\textwidth]{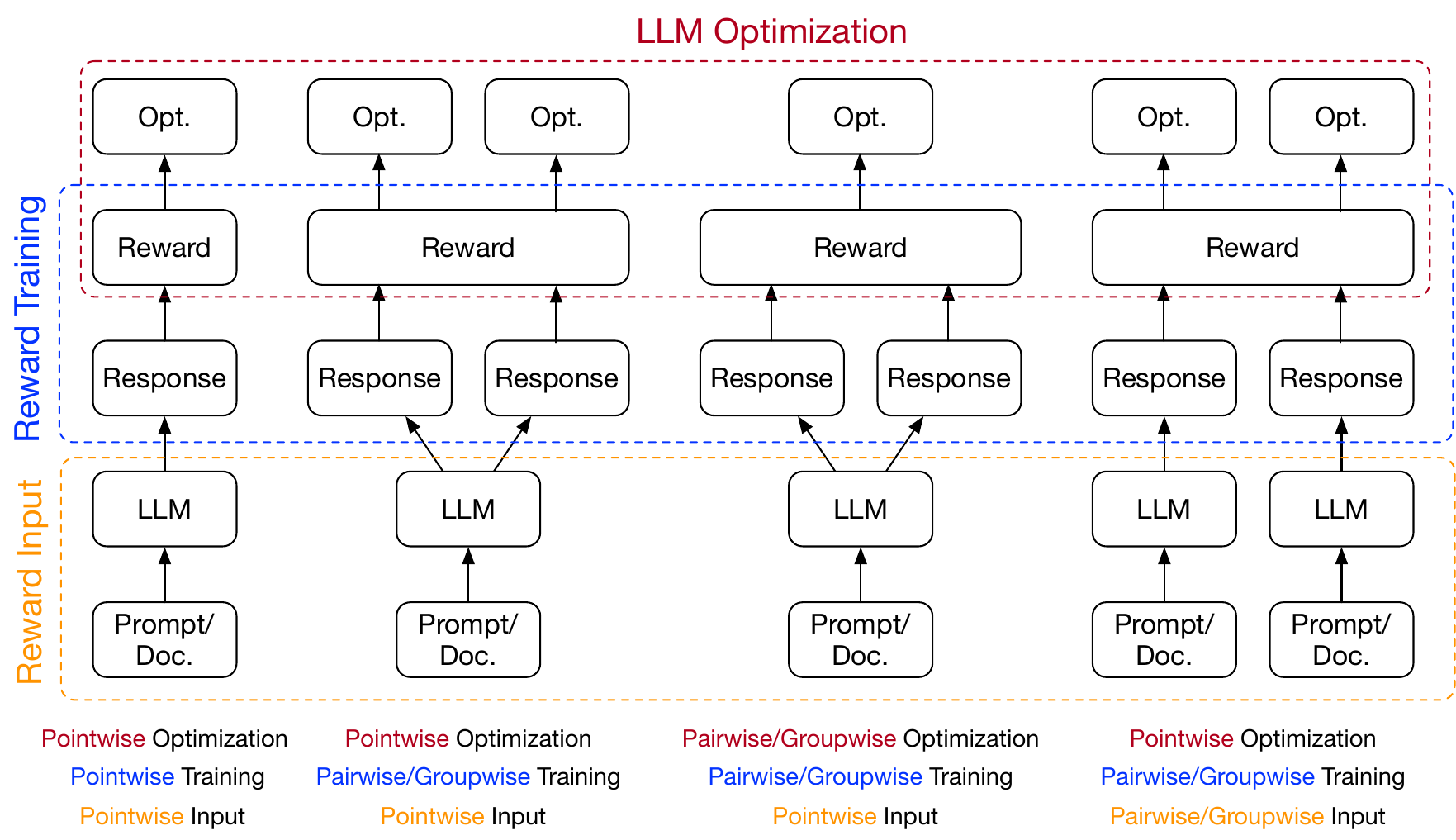} 
	\caption{An illustration of example reward collection and optimization methods in LLM alignment.}
	\label{fig:alignment_paradigm}
\end{figure}

The collection of rewards is a pivotal step in aligning LLMs. 
It evaluates the model's outputs/responses against certain criteria to determine how well they align with desired outcomes. Very much similar to the design of loss functions in learning to rank (LTR)~\cite{INR-016}, the nature of reward computations in LLM alignment can be broadly grouped into several categories based on the inputs and training paradigms of the reward functions. 
Specifically, if we borrow the terminology used in LTR literature~\cite{10.1145/3341981.3344218}, the reward collection methods can be categorized from two perspectives. 
From the perspective of reward function input, we have 
\begin{itemize}
	\item Pointwise Input: Rewards are computed for LLM outputs based on each individual input data points independently. 
	\item Groupwise (Pairwise) Input: Rewards are computed for LLM outputs based on a group (or pair) of different input data points together.
\end{itemize}
From the perspective of reward computation or reward function training, we have 
\begin{itemize}
	\item Pointwise Training: Rewards are computed on or reward functiosn are trained with each LLM output independently. 
	\item Groupwise (Pairwise) Training: Rewards are computed on or reward functions are trained with a group (or pair) of LLM outputs together.
\end{itemize}
An illustration of the differences between those methods is depicted in Figure~\ref{fig:alignment_paradigm}. 
With this taxonomy, we introduce a couple of popular alignment methods in the followings and summarize their types in Table~\ref{tab:alignment_types}. 
Careful readers may notice that all the reward methods here has a prefix ``RL'', which stands for \textit{Reinforcement Learning}. 
While these reward collection methods are independent to the use of learning algorithms (which is discussed in Section~\ref{sec:optimization}), they are often referred to or analyzed together with reinforcement learning.
For simplicity, we use the terminology widely used in the LLM literature to refer to them, but please note that this doesn't indicate that alignment methods using these rewards must be developed under reinforcement learning frameworks.

\textbf{Reinforcement Learning from Human Feedback (RLHF)}.
To the best of our knowledge, RLHF is the most well-known and popular alignment method today. 
It has been recognized as one of the most important parts of ChatGPT~\cite{openai2023gpt4}.  
RLHF collects feedback from human users on different LLM outputs to optimize model parameters accordingly~\cite{stiennon2022learning,NEURIPS2023_democrate}. 
Since the feedback is directly collected from human, RLHF is capable of aligning generative AI models with almost all related objectives such as output safety and ethical values. 
Typical RLHF process involves the generation of multiple output candidates from one or multiple LLMs given a single input.
It can be treated as a pointwise input method because RLHF always collect rewards on output candidates generated for a single input (e.g., prompt).
Then, with the output candidates generated by LLMs, RLHF further asks human annotators to judge the quality of each output and train reward functions accordingly.
The annotation process could be pointwise (through not common in recent LLM literature), i.e., asking the annotator to give a rating to each output separately, or pairwise/groupwise, i.e., asking the annotator to provide a preference or ranking of multiple outputs together.
Accordingly, the final reward function learned from such annotation data is constructed with either pointwise, pairwise, or groupwise training data, and thus can be categorized as pointwise, pairwise, or groupwise output methods.

\textbf{Reinforcement Learning from AI Feedback (RLAIF)}.
Despite the flexibility and effectiveness of RLHF, its needs for human in the loops significantly raise the cost of model alignment and unpredictable variance, particularly when the feedback data are not large or reliable enough, in the optimization process.
Therefore, researchers have also investigated extensively on how to conduct model alignment without supervised data.
Given the rapid development of LLMs in the last two years, one of the trending alignment methods in both academic and industrial communities is RLAIF~\cite{lee2023rlaif}.
The motivation of RLAIF is to replace human in the process of RLHF with a powerful LLM that can mimic human behaviors (which we refer to as the AI feedback model) to generate feedback for each model output.
Based on this idea, it basically reuses the existing framework of RLHF with minor modifications and align LLMs for different objectives by prompting the AI feedback models with objevtive-related task descriptions or annotation guidelines (e.g., RLCD\cite{yang2023rlcd}).
RLAIF is a pointwise input method and, theoretically, could be either pointwise, pairwise, or groupwise from the output perspective.
However, as far as we know, none existing studies have used RLAIF with the groupwise output paradigm, probably because directly ranking multiple candidates is still a difficult task for modern LLMs~\cite{sun2023chatgpt}.

\textbf{Reinforcement Learning from Contrastive Feedback (RLCF)}.
RLHF and RLAIF are powerful methods that have already been shown to be effective in many NLP tasks, but their applications to optimize alignment objectives specifically important in information access, unfortunately, have been unsuccessful so far.
As discussed in Section~\ref{sec:align_objective}, the ability to discriminate information in fine grains is the key to generate informative and useful outputs in IA scenarios, but studies have found that naively adopting those alignment techniques do not improve the model's performance in IA tasks as we expected~\cite{yoon2024ask,dong2023aligning}.
One of the key reasons lays in the input paradigms of RLHF and RLAIF. 
To generate outputs that are informative and discriminative, LLMs need to understand and capture what make an input piece of information unique in the corpus or data collections.
However, RLHF and RLAIF are developed with the pointwise input paradigm, and it's difficult, if not impossible, to teach LLMs to generate outputs unique to an input without seeing and comparing with other candidate inputs.
Therefore, RLCF is proposed to conduct alignment with groupwise input and output paradigms for IR~\cite{dong2023aligning}.
The idea is to let LLMs generate outputs for different inputs simultaneously and construct reward functions based on the comparison of each output for each input.
For example, one can compare query generation or expansion candidates for a single document with those generated for other similar documents to improve LLMs' ability in capture the uniqueness of each document.
The original RLCF method compute rewards with retrieval models to enhance the final LLM's effectiveness in IR tasks, but such groupwise input and output paradigms could have potentials in the alignment of other objectives as well because, as widely acknowledged in LTR literature, groupwise methods have more capacity in complex objective modeling and less variance in parameter optimization.




\subsubsection{Parameter Optimization}\label{sec:optimization}

With the rewards collected for particular alignment objectives, the next step is to optimize the parameters of LLMs.
Similar to the methods we used for feedback collection, the optimization algorithms for LLM alignments can also be broadly categorized based on their inputs. 
In this section, we only describe several representative optimization algorithms for LLM alignments, namely Proximal Policy Optimization (PPO)~\cite{schulman2017proximal}, Direct Preference Optimization (DPO)~\cite{NEURIPS2023_DPO}, and ranking-based optimization~\cite{yuan2023rrhf,dong2023raft}.
Please note that this is still an ongoing research direction, and the methods discussed here are far from covering all potential solutions in the area.

\textbf{Proximal Policy Optimization (PPO)}.
While PPO is not the first optimization algorithm used for reinforcement learning in LLMs, it is considered one of the most popular methods in LLM literature today, partially thanks to its applications in OpenAI products\footnote{\url{https://openai.com/research/openai-baselines-ppo}} and ChatGPT\cite{openai2023gpt4}. 
The primary goal of PPO, so as reinforcement learning techniques in general, is to train models to make sequences of decisions by rewarding desired behaviors and penalizing undesired ones. 
In the context of LLM alignment, PPO can be used with different types of rewards discussed in Section~\ref{sec:reward}, and it stands out from other RL algorithms due to its better balance of simplicity, efficiency, and effectiveness, compared to its predecessor such as Trust Region Policy Optimization (TRPO)~\cite{schulman2017trust}.
The core idea of PPO is to take small steps in policy space to improve the model while ensuring that the new policy is not too different from the old one. 
This is achieved through a clip mechanism, which limits the size of the policy update at each iteration. 
The clipped objective function helps the model conduct gradient descent while preventing overly large updates that could lead to performance collapse, a common issue in earlier RL methods that could lead to unstable training processes.
To compute such objective functions, PPO needs a reward model that can directly estimate the gain or loss of a particular action (e.g., the output of LLMs).
Therefore, it's usually used with pointwise output reward collection methods such as those shown in Table~\ref{tab:alignment_types}.
RLHF with PPO is widely used as the backbone alignment method of many famous LLMs such as GPTs, Llamas, etc. 

\textbf{Direct Preference Optimization (DPO)}.
A typical alignment method using PPO needs to create a reward model from the collected feedback data to score each LLM output for parameter optimization.
While the construction of such reward model is possible for most types of rewards, it doesn't necessarily fit the nature characteristics of each reward type.
It is essentially a pointwise output method that creates independent labels for each output candidate, and converting pairwise or groupwise data (e.g., human preferences over different LLM outputs) to pointwise format usually lead to significant information loss and more variance in model optimization~\cite{chu2024pre}.
To this end, Direct Preference Optimization (DPO)~\cite{NEURIPS2023_DPO} is proposed to optimize model parameters directly with preference data.
Careful readers may notice that all reward collection methods discussed in Section~\ref{sec:reward} has a prefix ``RL''. 
This is partially because most popular alignment methods use reinforcement learning for parameter optimization.
In contrast, DPO directly computes model gradients without using reinforcement learning by minimizing the KL divergence between the ground truth preference and LLM output distributions.
This method is highly similar to standard pairwise or listwise methods used in learning-to-rank literature~\cite{burges2005learning,10.1145/3209978.3209985,bruch2019analysis}. 
As pointed out by \citet{NEURIPS2023_DPO}, it outperforms popular reinforcement learning methods based on PPO in both effectiveness and robustness.
Considering that most alignment objectives (e.g., harmfulness, ethic, etc.) involve significant human subjectivity, preference-based optimization methods could be more promising in theory.
Besides, from the research perspective of information access, this also indicates that techniques from classic retrieval and ranking studies may provide important guidelines for the design of future LLM alignment methods.  

\textbf{Ranking-based Optimization}.
Following similar motivations with DPO, several methods have been proposed to further extend the utilization of pairwise preference reward to listwise reward for LLM alignments.
Notable representatives include RRHF\cite{yuan2023rrhf} and RAFT\cite{dong2023raft}. 
Despite data processing and implementation details, RRHF could be treated as a listwise version of DPO.
Its core idea is to score multiple responses via a crafted probability function and learns to align the corresponding probability distribution with human preferences through a ranking loss constructed based on the variation of hinge functions~\cite{liu2022brio}.
RAFT approaches the problem from a different angle.
It ranks multiple LLM response candidates based on preference data (or a reward model learned from preference data), and select samples with highest rewards to fine-tune the LLM.
It's well-acknowledged in IR literature that listwise ranking methods have better potentials in fitting preference data, both in theory and in practice~\cite{INR-016}.
Therefore, methods like RRHF and RAFT have both theoretical and empirical advantages over standard PPO and DPO methods in model alignments.
While such advantages are not fully explored in the general tasks such as dialog generation and machine translation, they could be important for the application of generative models in information access since many IA tasks exhibit natural needs of response discrimination and ranking.

 




\section{Learning from User Feedback in GenIR }\label{sec:userfeedback}

As introduced in the previous section, in the training procedure of LLMs, user feedback is very important to align the values of LLMs with humans. Reinforcement Learning from Human Feedback (\textbf{RLHF}) is widely adopted as the final training stage of LLMs. Besides LLMs, user feedback is vital in information retrieval systems. It is commonly used as the final optimization target. For example, the CTR task~\cite{chuklin2022click,zhou18ctr,gu2021ad} aims to predict the click-through rate, and the ranking models are usually tuned toward the user click signals~\cite{dou08click}. Besides, some user-centric tasks such as recommendation and personalized search collect and utilize user history feedback to provide tailored results for users' current information needs~\cite{liu10newsrec,isinkaye2015recommendation,qiu2006automatic}. For example, personalized search models apply the query attention technique to aggregate user search and click histories to build user preferences under their current queries~\cite{ge2018per}.

In the era of LLMs, many personalized search and recommendation~\cite{wu2024surveylargelanguagemodels} approaches devise LLMs to understand user histories and construct user interests. For example, inspired by the memorization mechanism in cognitive science, ~\citet{zhou_per} designed several memory modules including sensory memory, short-term memory, and long-term memory to facilitate LLMs to retrieve relevant user histories to current intents. In the recommendation area, LLMs are usually adopted to enrich user histories since they store extensive world knowledge~\cite{wu2024surveylargelanguagemodels}. Recently, LLM-based agents~\cite{agentsurvey} have attracted much attention from academia and industry. These agents have abilities to memorize past behaviors, make plans to achieve final tasks, and take action under current situations. It is worth exploring to involve user feedback in search agents to solve IR tasks.

\subsection{Continual Learning}
IR systems are designed to retrieve relevant information based on user queries. As users interact with these systems, they generate valuable data, such as search queries, clicked results, dwell time on pages, and explicit feedback like ratings or comments, that can be used to improve the system's understanding of user intent and preference. By leveraging the collected data, IR systems can progressively refine their retrieval algorithms, leading to more accurate and personalized search results. This process forms the basic paradigm of continual learning~\cite{wang2024comprehensive} in an IR system. Many methods have been proposed to incorporate user feedback data into optimizing a traditional information retrieval system~\cite{chapelle2009dynamic,dou2008click}. 

Continual learning is also vital for generative systems like LLMs to be regularly updated to include the latest human knowledge and feedback~\cite{wu2024continuallearninglargelanguage,shi2024continual}. As introduced by~\citet{wu2024continuallearninglargelanguage}, continual learning could be applied with different training stages, including pre-training, fine-tuning, and alignment. The traditional IR ranking models are relatively small and can be easily updated in a batch manner. The separated document index could be updated dynamically when new documents are available and hence it is relatively easier for the entire system to update continually. Contrarily, generative IR models are usually large and all information about the documents and the ranking are embedded in the same generative model. It is much more challenging to update such systems. For example, LLMs have the ``catastrophic forgetting'' problem~\cite{shi2024continual}: the performance of the old task based on previous knowledge domains will degrade when new user data are fed. 

\subsection{Learning and Ranking in Conversation Context}
In the interaction with conversational search systems, users may generate various types of feedback, such as asking follow-up questions based on the system's responses, expressing dissatisfaction with the system's responses, and providing clarification to the system's inquiries. These natural language-based explicit user feedbacks are crucial for helping the conversational search system continuously meet user needs and optimize its performance. LLMs possess powerful capabilities for understanding and generating dialogue, offering significant opportunities for better comprehension of user feedback in conversational search. 

In conversational search, the user questions are usually ambiguous and can only be correctly understood based on the conversation context.
Traditional methods are struggled in dealing with the long and complex conversation context, resulting in unsatisfactory retrieval performance.
In contrast, LLMs show outstanding capability in conversation understanding and therefore can largely improve the accuracy of conversational search intent understanding.
Mao et al.~\cite{emnlp23_llm4cs} proposed a prompting framework to leverage LLMs to perform conversational query rewriting. They developed three aggregation methods to aggregate the generated rewrites and hypothetical responses from LLMs to form a better search intent representation for conversational search.
Similarly, Ye et al.~\cite{emnlp23_EDQR} also proposed to utilize LLMs to generate informative query rewrites through well-designed instructions. 
Their results showed that the search performance can be largely improved after utilizing the generated contents from LLMs.
Furthermore, LLMs can also be used to mimic the users' search behaviors and generate more high-quality search session data.
Conversational search systems need massive session-level relevance data for improvements and LLMs can significantly facilitate the data curation process.
One of such related works is ConvAug~\cite{arxiv24_convaug}, which is a cognition-based framework that leverages LLMs to generate more conversational search sessions. These pseudo sessions can help conversational retrievers capture the diverse
nature of conversational contexts to be more effective and robust.

Besides, in the interaction process of conversational search, the user's responses to the system responses (e.g., clarification questions and inaccurate responses) are also crucial for capturing the users' real information needs and unique preferences.
Recently, TREC organized an interactive knowledge assistance (iKAT) track~\cite{trec_ikat} for studying collaborative conversational information-seeking systems that can customize and personalize their response based on what they learn about and from the user.
Existing works~\cite{arxiv23_llmrec_survey} have demonstrated the strong performance of LLMs in aggregating and inferencing users' references.
Therefore, LLMs have a large potential to improve the utilization of this type of valuable user initiative feedback to model the user profiles and provide a more accurate and personalized search experience.
LLMs can also be employed to identify the type of users' responses, such as distinguishing whether the response is a new question, a reply to a clarification request, or a hint for correcting a previous answer.
We do not need to train a separate model for this intent identification. Instead, we can stream the modeling of all interaction processes in conversational search through prompting with LLMs.
The massive knowledge about conversation patterns and the world of LLMs also makes it a promising end-to-end foundation to be an end-to-end foundation model for personalized conversational search systems.

\subsection{Prompt Learning}

LLMs have demonstrated excellent performance in language understanding, making them also promising for learning user feedback, particularly in the area of query refinement. In search engines and similar platforms, understanding the context and intent behind user queries is crucial for delivering accurate and relevant results. We consider two possible ways of applying LLMs to query refinement: 

\textbf{Directly Prompting LLMs for Query Refinement}. 
Given the substantial computational resources required for fine-tuning LLMs, a more straightforward approach is prompt learning. This method entails describing the task in text and prompting the models to solve it. Upon gathering user feedback, LLMs can analyze the feedback, comprehend the underlying meaning, and suggest refinements for the user input query, thereby enhancing retrieval performance. Previous studies~\cite{arxiv22_HyDE,ICLR23_GenRead,NIPS22_chain_of_thought} have applied LLMs to query rewriting. The results indicate that LLMs can generate effective user queries, particularly when provided with few-shot demonstrations. Furthermore, LLMs have shown superior performance in conversational query rewriting~\cite{emnlp23_llm4cs}, attributable to the availability of more comprehensive contextual information. These findings indicate the significant potential of applying LLMs to query refinement. 

\textbf{Distilling Knowledge from LLMs to Smaller Models.} 
In practice, it is still costly to use LLMs in real applications. Under this circumstance, training a small model specifically for query refinement emerges as a more favorable approach. This can be achieved by employing LLMs to refine queries based on user feedback, subsequently utilizing these refined queries as labels to train a specialized model. This strategy not only reduces computational overhead but also maintains the efficacy of the learning process, thereby offering a pragmatic solution for real-world applications.

\section{Summary} \label{summary}
In this chapter, we delve into how user feedback can enhance the GenIR system. Firstly, we clarify the concept of "user" and subsequently explore the diverse types and forms of user feedback information. Furthermore, we outline four established strategies that leverage user feedback effectively. Secondly, we provide a detailed account of the crucial technique of alignment in the GenIR context, discussing both the alignment objective and various methods employed. Finally, we highlight the significance of user feedback learning in GenIR, encompassing human-in-the-loop approaches, continuous learning, learning and ranking within conversational contexts, as well as prompt learning. Through this comprehensive exploration, it becomes evident that innovative techniques are being proposed beyond traditional methods of utilizing user feedback, and contribute significantly to the evolution of GenIR in the new era.

There are some challenging topics and future directions that we believe need further exploration, such as:

\begin{itemize}
    \item \textit{User intention understanding within the GenIR system}. For example: How do we precisely determine the user’s true intent? How do we manage shifts in user intentions during multi-turn interactions or conversations with the GenIR system? When we broaden the concept of \textbf{\textit{user}} to also include agents/clients that interact with the GenIR system, could this lead to self-feedback loops within the GenIR system and a bias towards artificial intentions? 

    \item  \textit{User behavior analyzing and understanding with "less but rich feedback"}. As the end user interact with generated responses, we may receive less feedback than traditional IR systems (e.g., clicks on search engine results pages). On the other hand, the feedback is richer (e.g., an explicit feedback in the conversation like “thank you, that's really helpful” or a detailed follow-up indicating continued engagement when the information need is not met). Studying the utilization of limited yet in-depth user interaction behaviors in the GenIR system is valuable. There are additional research questions, such as: how do we align personalized models using limited user data? How can we efficiently fine-tune and store personalized generative models? 

    \item  \textit{User-centric evaluation of the GenIR system}. For instance, how do we measure user satisfaction when engaging with complex tasks during interactions with the GenIR system? Is personalized evaluation feasible and essential?    

    \item  \textit{Privacy protection within the GenIR system}. Particularly, we need to consider how to ensure privacy is maintained when utilizing user feedback in personalized generative models.
\end{itemize}


\bibliography{sn-bibliography}

\begin{thebibliography}{84}
\bibitem[\protect\citeauthoryear{Vincent}{}]{Vincent2016}
\begin{botherref}
\oauthor{\bsnm{Vincent}, \binits{J.}}:
Twitter taught microsoft’s ai chatbot to be a racist asshole in less than a
  day
\end{botherref}
\endbibitem
\end{thebibliography}

\end{document}